\documentstyle [12pt,epsf] {article} 
\begin{document}
\begin{titlepage}
\begin{center}
\vspace{2cm}
\LARGE
The K-band luminosity function at $z=1$: a powerful constraint on galaxy formation theory\\
\vspace{1cm} 
\large
Guinevere Kauffmann$^{1}$ \& St\'{e}phane Charlot$^{2}$ \\
\vspace{0.5cm}
\small
{\em $^1$Max-Planck Institut f\"{u}r Astrophysik, D-85740 Garching, Germany} \\
{\em $^2$ Institut d'Astrophysique du CNRS, 98 bis Boulevard Arago, F-75014 Paris, France} \\
\vspace{0.8cm}
\end{center}
\normalsize
\begin {abstract}
There are two major approaches to modelling galaxy evolution.
The  ``traditional'' view is that the most massive galaxies were assembled early
and have  evolved with steeply declining star formation rates since a 
redshift of 2 or higher. 
According  to hierarchical theories, massive galaxies were assembled much more 
recently from mergers of smaller subunits.
Here we present a simple observational test designed to differentiate
between the two. 
The observed K-band flux from a galaxy is a good
measure of its stellar mass even at high redshift. It is likely only weakly
affected by dust extinction.
We compute the evolution of the observed K-band luminosity function for traditional,
pure luminosity evolution (PLE)
models and  for hierarchical models. 
At $z=0$, both models can fit the observed local K-band luminosity function.
By redshift 1, they differ greatly in the predicted abundance of bright galaxies.
We calculate the redshift distributions of K-band selected galaxies and compare
these with available data. We show that the number of 
$K < 19$ galaxies with redshifts greater than 1 is well below the numbers predicted
by the PLE models. In the Songaila et al (1994) redshift sample of 118 galaxies with $16<K<18$,
33 galaxies are predicted to lie at $z>1$. Only 2 are observed. In the Cowie et al (1996) 
redshift sample of 52 galaxies with $18<K<19$, 28 galaxies are predicted to lie at $z > 1$. Only 5 are observed.
Both these samples are more than 90\% complete.
We conclude that there is already strong evidence that the abundance of massive galaxies
at $z \sim 1$ is well below the local value. This is inconsistent with the traditional
model (unless most massive galaxies are extremely heavily obscured by dust             
at redshift 1), but similar to the expectations of hierarchical models.

\end {abstract}
\vspace {0.8 cm}
\end {titlepage}

\section {Introduction}          
At present there are two main approaches to modelling galaxy evolution. 
The ``traditional'' view
is that the most massive galaxies -- the big ellipticals, S0s and early-type spirals -- formed first
and have been evolving with strongly declining star formation rates since at least a  
redshift of 2. The less massive late-type spiral and irregular galaxies
formed their stars at a roughly constant rate over a Hubble time and in many cases undergo
substantial bursts. Alternatively,
according  to hierarchical theories of galaxy formation, massive galaxies were assembled 
recently from mergers of smaller subunits.

Traditional models adopt a simple ``backwards-in-time'' technique
for calculating how the observed properties of galaxies evolve as a function of
redshift (e.g. Tinsley 1980; Bruzual \& Kron 1980; Koo 1981; Shanks et al 1984;
King \& Ellis 1985; Yoshii \& Takahara 1988; Guiderdoni \& Rocca-Volmerange 1990).
The two basic inputs are a) the observed present-day luminosity
function divided by morphological type, and b) a parametrization
of the mean  star formation rate history of galaxies of a given morphological type. These star
formation histories are tuned to reproduce the spectral energy distributions
of nearby spirals and ellipticals. It is usually assumed that  all galaxies
form at the same redshift and evolve as  closed
box systems. Stellar population synthesis models are used to compute how the galaxy
luminosity function evolves as a function of
lookback time, and predictions are then made for the counts, 
redshift distributions and colours of faint galaxies.

Hierarchical models of galaxy formation follow the formation and evolution of galaxies
within a merging hierarchy of dark matter halos (e.g. White \& Rees 1978; White \& Frenk 1991;
Lacey et al 1993; Kauffmann, White \& Guiderdoni 1993; Cole et al. 1994; Somerville 1997). 
An analytic formalism allows 
the progenitors of a present-day object, such as a cluster, to be traced back
to arbitrarily early times. Simple prescriptions                                   
are adopted to describe gas cooling, star formation, supernova feedback and the
merging of galaxies. Finally, stellar population
synthesis models are used to generate luminosity functions, counts and redshift
distributions for comparison with observations.

There have been many recent papers claiming that either the traditional
models or the hierarchical models can provide good fits to available data on high
redshift galaxies. Most of the recent papers discussing traditional models invoke
a local luminosity function with a ``steep'' ($\alpha < -1$) faint-end slope
in order to reproduce the observed high number density of faint blue galaxies.
Dust extinction is also included to avoid overproducing  
blue, star-forming galaxies at high redshift (Gronwall \& Koo 1995; Campos \& Shanks 1997;
Pozzetti et al 1996).
The hierarchical models {\em predict} a local luminosity function with a steep
faint-end slope (perhaps too steep) and hence have little trouble in matching the counts. Because
bright galaxies have not yet formed at high redshift, the hierarchical models
can reproduce the redshift distributions without requiring dust 
(Kauffmann, Guiderdoni \& White 1994;
Cole et al 1994; Heyl et al 1995; Baugh, Cole \& Frenk 1996).

In this Letter, we present an observational test that is designed to {\em differentiate}
between the traditional and hierarchical scenarios. To do this, we concentrate
on the aspect that differs the most between the two pictures 
-- the redshift evolution
of the most massive galaxies. We show that the observed K-band flux from a galaxy provides a good
measurement of its total stellar mass out to redshifts in excess of 2 for all but the most
extreme starbursting systems. Moreover, the K-band 
is only weakly affected by extinction at these redshifts.
 We compute the evolution of the observed K-band
luminosity function for  traditional and hierarchical models, assuming both 
high-density ($\Omega=1$)
and  low-density ($\Omega=0.2$) cosmologies. We use the new population synthesis models
of Bruzual and Charlot (in preparation), which include updated stellar evolutionary
tracks and new spectral libraries.
At $z=0$, both the traditional models and the $\Omega=1$ hierarchical model can give a good
fit to the local  K-band luminosity functions of Gardner et al (1997) and
Szokoly et al (1998). By redshift 1, the two
 give very different predictions for bright galaxies.
At $z=1$, the traditional models predict 10 times more galaxies with $K<17$
than do the hierarchical models.
At $z=2$, the traditional models
predict fifty times more galaxies with $K < 19$.

We then calculate the redshift distributions of galaxies with $K < 19$ and compare
these with the Hawaii Deep Field Samples of Songaila et al (1994) and 
Cowie et al. (1996).   We show that the observed number of
bright galaxies with redshifts greater than 1 is well below the number predicted
by the traditional models. In the Songaila et al (1994) sample of 118 galaxies with $16 < K < 18$,
33 galaxies are predicted to lie at $z>1$. Only 2 are observed. In the Cowie et al (1996)
sample of 52 galaxies with $18 < K < 19$, 28 are predicted to lie at $z > 1$. Only 5 are observed.
Both these samples are more than 90\% complete.

We conclude that there is already strong evidence that the abundance of massive galaxies
has declined substantially at $z \sim 1$. This is inconsistent with the traditional models, unless 
most massive galaxies are extremely heavily obscured by dust at this redshift, but similar to the
expectations of hierarchical models.

\section {Relating observed K-magnitudes to stellar mass}
In figure 1, we show the observed K-magnitude of a galaxy with $10^{11} M_{\odot}$ of stars
as a function of its  observed redshift. We assume $q_0=0.5$, $H_0= 50$
km s $^{-1}$ Mpc$^{-1}$. (Note 
that this is {\em not} an evolutionary
plot for any specific galaxy. Rather it shows the apparent brightness of a galaxy which has          
$10^{11} M_{\odot}$ of stars {\em at the redshift when it is observed.})
The three different curves in the plot are for three
possible star formation histories of the galaxy. 
The solid line shows the predicted K-band magnitude   
of the galaxy if all its stars formed in a burst at $z= \infty$. The short-dashed line
shows the K-magnitude if the galaxy formed its stars at a constant rate from $z=\infty$ to the
redshift of observation.
The long-dashed line is for a galaxy that formed 80\% of its stars at a constant
rate, and 20\% over the $10^{8}$ years immediately before it is observed. 
This is supposed to represent a
galaxy that is undergoing a strong starburst. In the lower panel of figure 1, we plot
the observed $B-K$ colours for the same three star formation models.
Note that we have adopted a Scalo IMF with upper and lower mass cutoffs of 100 $M_{\odot}$
and 0.1 $M_{\odot}$. A 10 Gyr single-age stellar population has a K-band mass-to-light
ratio of $\sim 0.6$, in good agreement with the estimated K-band mass-to-light ratios
of elliptical galaxies (Mobasher et al 1998).

Figure 1 shows that galaxies of the same stellar mass will have roughly the same observed K-magnitude,
independent of their star formation histories. The difference in K-luminosity between
the initial burst and constant star formation models is less than a factor 2.
Even the starburst model differs from the initial burst model by less than a factor
3 at $z<2$. This is in marked contrast to the observed B-band luminosities,
which, for high redshift galaxies,
are dominated by the flux from young massive stars. Differences in $B-K$
colour of up to 9 magnitudes are obtained for different star formation models.
From figure 1, one can also infer  that evolution at $z<2$  in the abundances of galaxies
more massive than  $10^{11} M_{\odot}$ 
should be evident even in relatively bright ($17<K<19$) K-selected
samples.  

\begin{figure}
\centerline{
\epsfxsize=8cm \epsfbox{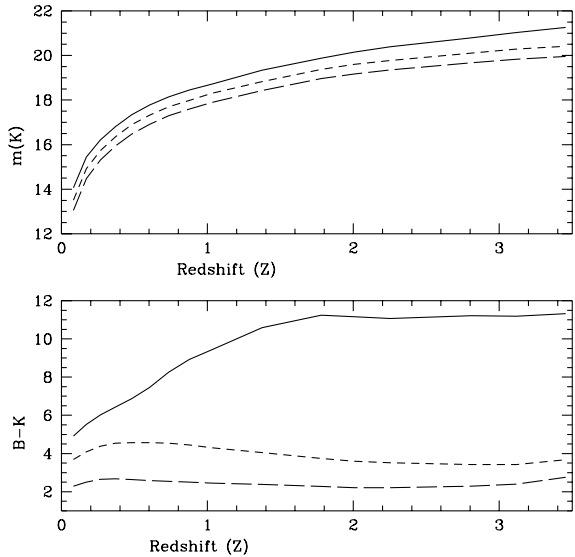}
}
\caption{\label{fig1}
\small
 {\em Upper panel:} The apparent K-band magnitude of a galaxy with
 $10^{11} M_{\odot}$ of stars observed
at redshift $z$. The solid line shows results if all stars form in a burst at $z=\infty$. The
short-dashed line assumes constant star formation until the epoch of
observation. The long-dashed line
is a ``starburst'' model (see text). {\em Lower panel:} The observed B-K colours
of these galaxies.}
\end {figure}

\normalsize

\section {The Models}
\subsection {The traditional models}
The traditional or pure luminosity evolution (PLE) model we employ is similar
to that described in Pozzetti et al. (1996). The present-day luminosity function is divided
into a set of 4 Hubble types (E/S0, Sa-Sb, Sc-Sd, Sm-Im) using the results of Marzke et al (1994)
on the relative abundances  of galaxies of differing morphology in the CfA Redshift Survey.
Because the Schechter-function $M^{*}$ derived for the CfA survey is considerably
fainter than that found for other recent surveys covering a larger area, we have chosen to
normalize the total $B$-band luminosity function using the Schechter function fits derived
for the ESO Slice Project (ESP) redshift survey ($\alpha=-1.22$, $M_{*}=-19.61+5 \log h$,
$\phi^{*}=0.020 h^{3}$ Mpc$^{-3}$) (Zucca et al 1997). At the bright end, the ESP luminosity function
agrees well with the luminosity functions derived for the Stromlo-APM and the 
Las Campanas redshift surveys (Loveday et al 1992; Lin et al 1996). We compute the spectral
evolution of E/S0, Sa-Sb and Sc-Sd galaxies using Bruzual \& Charlot (1998) models with
exponentially declining star formation rates with star formation timescales of 1, 4 and 15 Gyr, 
respectively. For the adopted Scalo (1986) initial mass function, solar metallicity and 
formation redshift $z_f=5$, the models provide good fits to the spectral features and the colours of
nearby galaxies of these types from the ultraviolet to the infrared (Table~1; Bruzual \&
Charlot 1998). Local Sm-Im galaxies are too blue to be fit by models 
with exponentially declining or even constant star formation rates, if we assume a formation redshift of 5.
Instead, we have adopted a model
with constant star formation rate seen at a fixed age of 1.4~Gyr. This model best reproduces the colours
and spectral properties of the propotypical Sm-Im galaxy NGC~4449 (Table 1; Bruzual \& Charlot 1998).
Since Sm-Im galaxies contribute very little to the luminosity function at bright
magnitudes, their treatment in the models does not affect our conclusions.       
Our results {\em are} affected by the evolutionary model we adopt 
for the luminous E/S0 and Sa-Sb galaxies. We note that the choice of                    
a Scalo IMF is a conservative one, 
since this results in  milder luminosity evolution than the Salpeter IMF,
which has a higher fraction of massive stars (Pozzetti et al 1996). We have not attempted
to include any metallicity evolution in the traditional models. This will also not affect our
conclusions since massive galaxies, which have short star formation timescales, quickly
reach solar metallicity.

\subsection {The hierarchical models}
Because of the simplified way in which they treat the many physical processes which play
a role in galaxy formation, hierarchical models have considerably more freedom than the
traditional models. On the other hand the properties of the present galaxy population, which
are used as an input in the traditional approach, are an output of the hierarchical models
and so can be used to test them. As examples of hierarchical models, we here use a model from
our earlier paper on the systematics of elliptical galaxies together with a low density
variant.
The reader is referred to  Kauffmann \& White (1993), Kauffmann, White \& Guiderdoni (1993)  
and Kauffmann \& Charlot (1998) for more details of our  semi-analytic techniques for
modelling the formation and evolution of galaxies in a hierarchical Universe. 
In the current paper we explore two possibilities:
\begin {enumerate}
\item An $\Omega=1$ ``standard'' CDM model with $H_0=50$ km s$^{-1}$ Mpc$^{-1}$, 
$\sigma_8=0.67$, and
 $\Gamma=0.5$. The star formation and feedback parameters are the same as in model A of
 Kauffmann \& Charlot (1998), which was shown to provide a good fit to the                   
 slope and scatter of the colour-magnitude relation of elliptical galaxies
 in clusters. This model includes a prescription for chemical evolution and makes use
 of the metallicity-dependent spectral synthesis models of Bruzual \& Charlot (1998).

\item An $\Omega=0.2$ ($\Lambda =0$) CDM model with $H_0=50$ km s$^{-1}$ Mpc $^{-1}$ 
 and $\sigma_8=1$. Chemical evolution is also included in this model.
\end {enumerate}

\subsection {The present-day $K$-band luminosity function}
Redshift zero $K$-band luminosity functions for the traditional and hierarchical models are shown
as thick lines in figure 2.  The Schechter function fit obtained by Szokoly et al (1998) for a sample
of 175 galaxies to $K$=16.5 is shown as a dotted line on the plot. The fit obtained by      
Gardner et al (1997) for a wide-angle survey of 567 galaxies limited at $K$=15,
is shown as a thin dashed-dotted line.
The traditional model and the $\Omega=1$ CDM model both fit the observed local $K$-band luminosity
function well, particularly at the bright end. 

Note that in the traditional model, the $K$-band
luminosity function is computed from the observed $B$-band luminosity function and the model $B-K$
colors corresponding to the different morphological types.  The fact that the K-band luminosity function
agrees well with the data gives us confidence that the 
the assumptions made in \S3.1 are giving self-consistent results.

The $\Omega=0.2$ CDM  model fails to produce enough galaxies at
$M^{*}$ by almost an order of magnitude. This failure of low-density models has been noted
before (Kauffmann, White \& Guiderdoni 1993). Baugh et al (1998) fit the number density of
galaxies at the ``knee'' of the luminosity function in a low-density ($\Omega=0.3, \Lambda=0.7$)
CDM model by adopting a brighter normalization, but they then vastly overproduce galaxies
at the bright end of the luminosity function. (Basically, this is equivalent to shifting
our curve to the right). Because of the failure of low-density hierarchical models to reproduce the
{\em shape} of the local $K$-band luminosity function, we will not pursue their predictions
any further in this Letter.

\begin{figure}
\centerline{
\epsfxsize=11cm \epsfbox{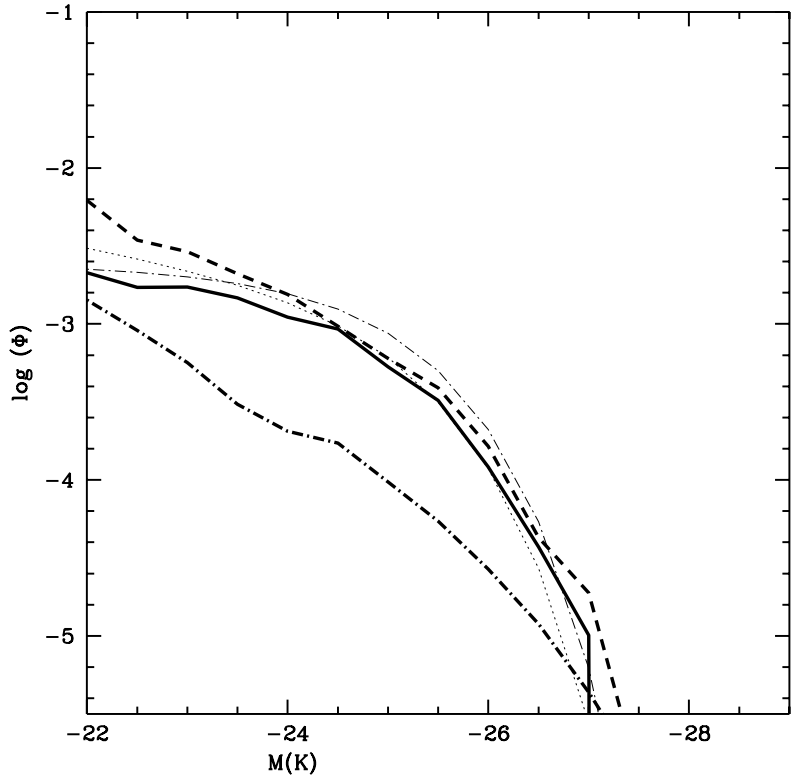}
}
\caption{\label{fig2}
\small
 The present-day $K$-band luminosity function of models (thick lines) compared with the 
 data (thin lines).               
The short-dashed line is the $\Omega=1$ hierarchical model, the dashed-dotted line
is the $\Omega=0.2$ hierarchical model and the 
solid line is the PLE model.                  
The dotted line is the Schechter fit serived by Szokoly et al (1998).
The thin dashed-dotted line is the fit obtained by Gardner et al (1997).}
\end {figure}

\normalsize
\subsection { Evolution of the $K$-Band luminosity function}
The predicted evolution of the observed differential
$K$-band luminosity function is shown in figure 3. The solid     
and dotted lines show results from the traditional models, for  $q_0=0.5$ and $q_0=0.1$ respectively.
The dashed line shows results from the $\Omega=1$ CDM model.
All models assume $H_0=50$ km s$^{-1}$ Mpc$^{-1}$.

By $z=1$, the predictions of the traditional and the hierarchical models deviate strongly at the 
bright end of the luminosity function. This just reflects the fact that in the hierarchical models,
the most massive galaxies are still forming today. In the traditional models, the massive galaxies
were already in place at $z=1$  and were 
significantly brighter than at present because their stars were less     
evolved.

\begin{figure}
\centerline{
\epsfxsize=13cm \epsfbox{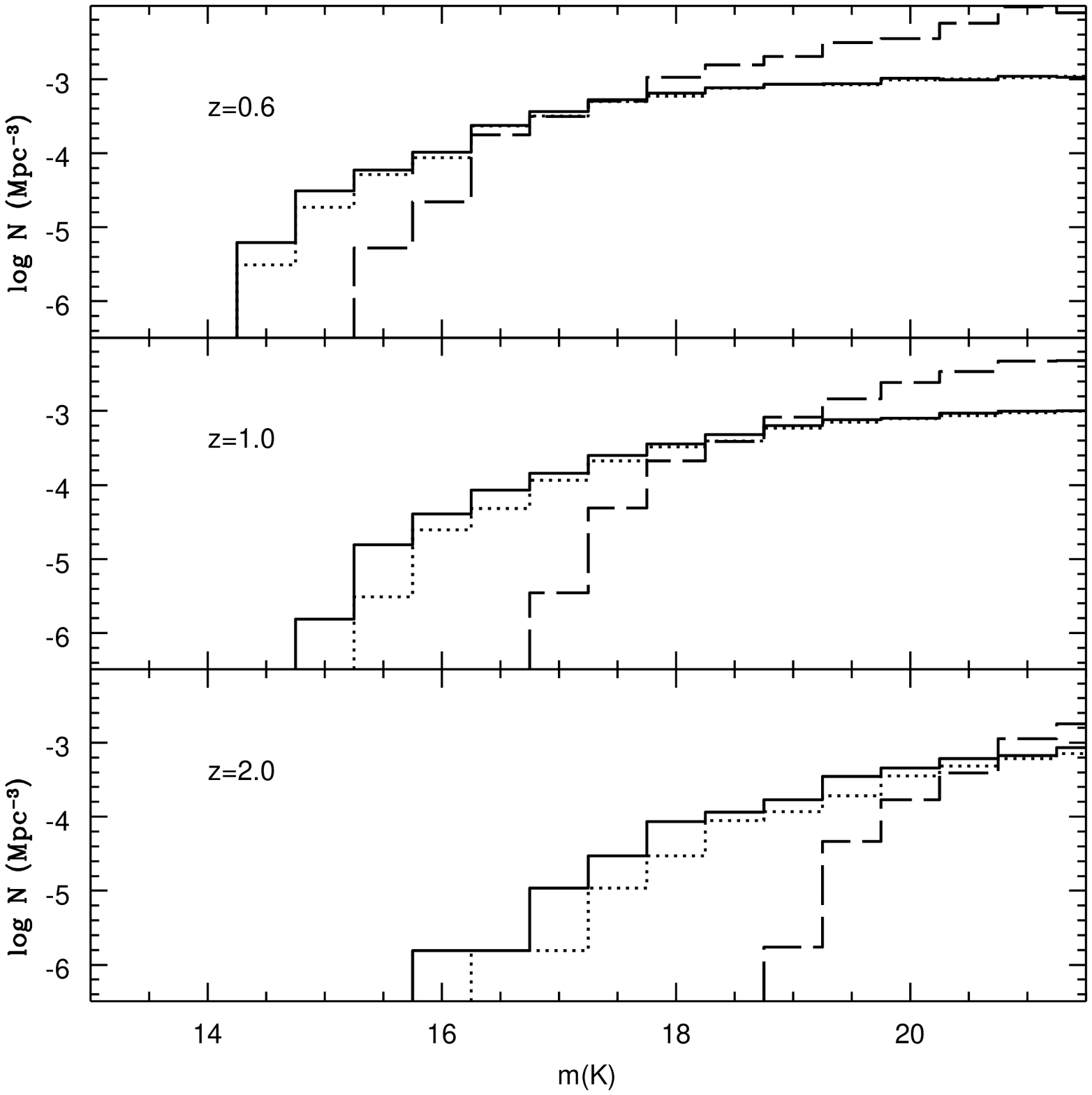}
}
\caption{\label{fig3}
The evolution of the observed differential $K$-band luminosity function of galaxies.
Solid and dotted lines show the $q_0=0.5$ and $q_0=0.1$ PLE models. The
dashed line is the hierarchical model.}
\end {figure}

\section {Comparison with data -- the redshift distributions of $K$-selected galaxies}

Complete $K$-selected redshift surveys of faint galaxies are not yet large enough
to construct accurate luminosity functions at a series of redshifts. The young
bright galaxies predicted by the traditional models should, however, show up as
a ``tail'' of high redshift galaxies in these samples. In the B-band, the presence
of high-z tails in the predicted redshift distributions has been shown to
depend sensitively on whether dust extinction is included in the modelling (Gronwall \& Koo 1995;
Campos \& Shanks 1997). In the $K$-band, the effects of dust are not likely to be as significant.

In figure 4, we plot the cumulative fraction of galaxies with redshifts greater than $z$ for samples 
limited at $K$=18, 19 and 21. The solid and dotted line show results for the $q_0=0.5$ and
$q_0=0.1$ PLE models. The dashed line is the  hierarchical model. 
The thin solid lines in figure 4 show recent observational results. The curve
in the top panel is computed using the sample of Songaila et al (1994) and the curve in
the middle panel is derived from the Hawaii Deep Field sample of Cowie et al (1996).
In both cases, we have chosen  conservative limiting magnitudes in order to ensure that
the redshift data are nearly complete ($>$ 90\% in both cases) and that the derived redshift
distributions are not biased by selection effects. There are 118 galaxies with $16 < K < 18$
in the Songaila redshift sample and 52 galaxies with $18 < K < 19$ in the Cowie redshift sample. 
In the plot, we
do not include galaxies without known redshifts (11 galaxies in Songaila et al. and
4 galaxies in Cowie et al.)

It should be noted that all the models produce roughly the right total number of galaxies
per unit area on the sky to $K=21$.
The $q_0=0.5$ PLE model fits the counts extremely well to $K$=21. The $q_0=0.1$ model counts are
a factor of two too high at $K$=21, because of the increased volume at high redshift
in a low-density cosmology. The hierarchical model counts are also a factor 2-3 too high at this 
magnitude,  because the faint end of the $K$-band luminosity 
function is too steep in this model (see figure 2). This accentuates the tendency for 
the hierarchical redshift distributions to be dominated by faint, low-z galaxies.

Figure 4 shows that the fraction of galaxies predicted to lie at high redshift is very much
larger in the PLE models than in the hierarchical model.
It is also  apparent that the number of galaxies observed at redshifts
greater than $z \simeq 0.5$ falls well below the predictions of the PLE models in both
the Songaila and the Cowie data samples. For  the $16<K<18$ sample,  the  PLE
models predict that 33 galaxies should have been detected at $z >1$, when in fact only
2 such galaxies are found.  
A KS test gives the probability  that the Songaila
sample is consistent with the PLE redshift distribution as less than 10$^{-4}$, even if all
the galaxies without redshifts are arbitrarily  assigned to $z>4$.
A similar result is obtained for the Cowie et al. sample. This shows that our conclusions are
not being affected by finite sample statistics.
In anticipation of future redshift surveys complete to even fainter limiting magnitudes, we show 
the predicted redshift distributions for galaxies with               
$19<K<21$ in the bottom panel of figure 4.

Finally, we note that similar results  have been found by others. Pozzetti et al (1996) show
that their PLE models fail to match the Songaila et al. (1994) redshift distributions.
Cowie et al (1996) note that the median redshift of their sample as a function of observed K magnitude
falls below the predictions of models with mild luminosity evolution.
In our analysis, we focus our attention on the evolution of the {\em bright end} of the K-band
luminosity function, since this where the predictions of the traditional
and the hierarchical models diverge most strongly.

\begin{figure}
\centerline{
\epsfxsize=13cm \epsfbox{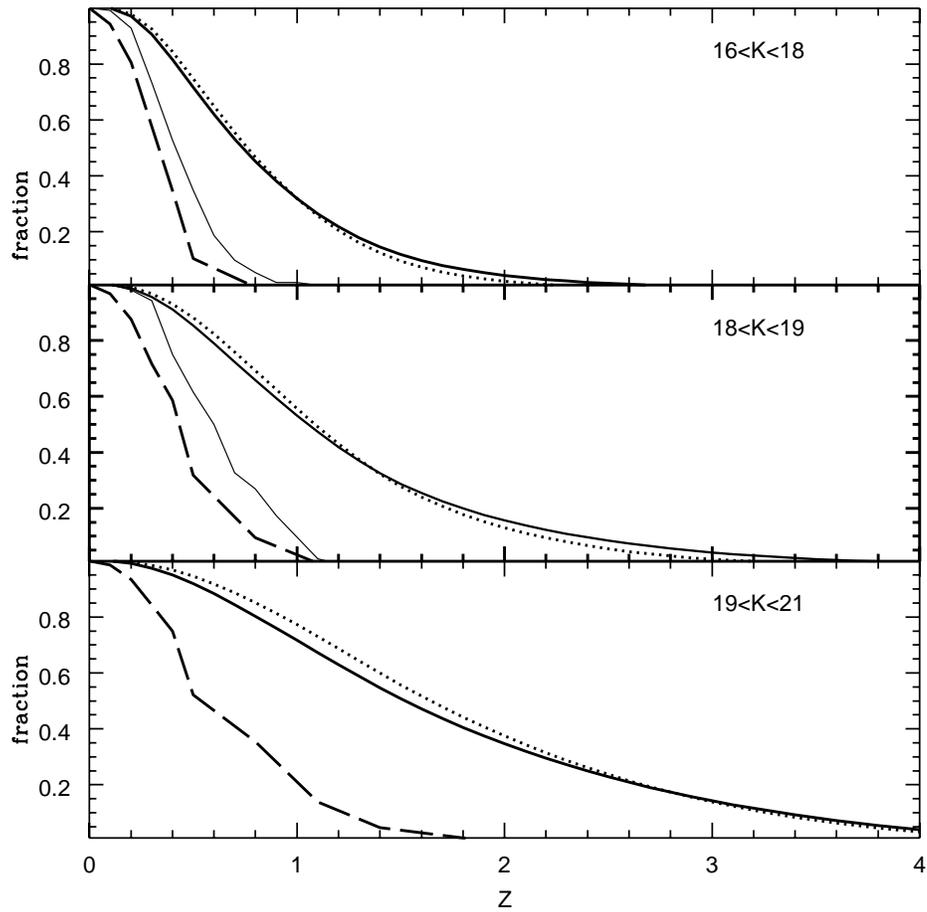}
}
\caption{\label{fig4}
The redshift distributions of galaxies selected according to $K$-magnitude.
Solid and dotted lines show the $q_0=0.5$ and $q_0=0.1$ PLE models. The
dashed line is the hierarchical model. The thin solid line is derived from
the Songaila et al (1994) sample in the upper panel and from the Cowie et al (1996)
sample in the middle panel.}
\end {figure}

\section {Discussion and Conclusions}

In previous work we showed that the abundance of galaxies with colours consistent with
passively evolving early-type galaxies  was a factor 2-3 lower  at a redshift of 1 than at present
(Kauffmann, Charlot \& White 1996; see also Zepf 1997). It was not clear from  our analysis
whether this was because present-day ellipticals                      
were forming stars at high redshift and were thus 
not as red as specified by simple passive evolution models, 
or because many ellipticals had simply not yet assembled by $z=1$.

The analysis presented in this Letter is considerably more general in that it applies to
the evolution of the {\em entire population of massive galaxies}. From our comparison of the data
with the models, we conclude that  the abundance of massive
galaxies is substantially below its present value at $z=1$, in contradiction with the
traditional picture of galaxy formation. If massive galaxies were forming stars more 
rapidly than we have assumed at high redshift,
they would appear even brighter at $K$, and the failure of the traditional picture
would be even more significant.
The only remaining ``escape route'' 
would be for a large fraction of today's massive galaxies to be so heavily obscured by dust
at $z=1$, that even the observed $K$-band is significantly affected (Franceschini et al 1997).
This does not seem likely to us, because
damped Lyman-alpha systems, which are good probes of the conditions within
high column density gas clouds, 
contain rather little dust at z $\simeq 2$  (Pei, Fall \& Bechtold 1991; Pettini et al. 1997)

We suggest that massive galaxies have continued forming until recent times  through a process 
of  merging and 
accretion, as predicted by hierarchical theories of galaxy formation. 
Future wide-area $K$-selected redshift surveys will enable this  buildup of galaxies to be
quantified more accurately. 

\vspace{0.8cm}

\large
{\bf Acknowledgments}\\
\normalsize
We thank Simon White for helpful discussions.                                
This work was carried out under the
auspices of EARA, a European Association for Research in Astronomy, and the TMR              
Network on Galaxy Formation and Evolution  funded by the European Commission.

\pagebreak

\vspace {1.5cm}
\normalsize
\parindent 7mm  
\parskip 8mm

{\bf Table 1:} Star formation laws and predicted colours of different Hubble types          
\vspace {0.3cm}

\begin {tabular} {llcc}
 Type & SFR & B-V (z=0) & V-K (z=0) \\                                                                  
 E/S0 & exp, $\tau=1$ & 0.97 & 3.22 \\
 Sa/Sb& exp, $\tau=4$ & 0.83 & 3.00 \\
 Sc/Sd& exp, $\tau=15$ & 0.61 & 2.75 \\
 Sm/Im& const, age=1.4 & 0.28 & 2.00 \\
\end {tabular}

\pagebreak

\Large
\begin {center} {\bf References} \\
\end {center}
\normalsize
\parindent -7mm  
\parskip 3mm

Baugh, C.M., Cole, S. \& Frenk, C.S., 1996, MNRAS, 283, 1361

Baugh, C.M., Cole, S., Frenk, C.S. \& Lacey, C., 1998, APJ, in press

Bruzual, G. \& Kron, R.G., 1980, APJ, 241, 25

Bruzual, A.G. \& Charlot, S., 1993, APJ, 405, 538

Campos, A. \& Shanks, T., 1997, MNRAS, 291, 383

Cole, S., Arag\'on-Salamanca, A., Frenk, C.S., Navarro, J.F. 
\& Zepf, S.E., 1994, MNRAS, 271, 781

Cowie, L.L., Songaila, A., Hu, E.M. \& Cohen, J.G., 1996, AJ, 112, 839

Franceschini, A., Silva, L., Granato, G.L., Bressan, A. \& Danese, L., 1997, APJ, submitted

Gardner, J.P., Sharples, R.M., Frenk, C.S. \& Carrasco, B.E., 1997, APJ, 480, 99

Gronwall, C. \& Koo, D.C., 1995, APJ, 440, L1

Guiderdoni, B. \& Rocca-Volmerange, B., 1990, A\&A, 227, 362

Heyl, J.S., Cole, S., Frenk, C.S. \& Navarro, J.F., 1995, MNRAS, 274, 755

Kauffmann, G. \& White, S.D.M. 1993, MNRAS, 261, 921

Kauffmann, G., White, S.D.M. \& Guiderdoni, B. 1993, MNRAS, 264, 201 (KWG) 

Kauffmann, G., Guiderdoni, B. \& White, S.D.M., 1994, MNRAS, 267, 981

Kauffmann, G., Charlot, S. \& White, S.D.M., 1996, MNRAS, 283, L117

Kauffmann, G. \& Charlot, S., 1998, MNRAS, in press

King, C.R. \& Ellis, R.S., 1985, APJ, 288, 456

Koo, D.C., 1981, PhD thesis, Univ. of California at Berkeley

Lacey, C., Guiderdoni, B., Rocca-Volmerange, B. \& Silk, J., 1993, APJ, 402, 15

Lin, H., Kirshner, R.P., Schechtman, S.A., Landy, S.D., Oemler, A., Tucker, D.L. 
\& Schechter, P.L., 1997, APJ, 464, 60
  
Loveday, J., Peterson, B.A., Efstathiou, G. \& Maddox, S.J., 1992, APJ, 390, 338 

Marzke, R.O., Geller, M.J., Huchra, J.P. \& Corwin, H.G., 1994, AJ, 108, 437

Mobasher, B., Guzman, R., Arag\'on-Salamanca, A. \& Zepf, S., 1997, MNRAS, in press  
    
Pei,Y.C., Fall, S.M. \& Bechtold, J., 1991, APJ, 378, 6

Pettini, M., King, D.L., Smith, L.J. \& Hunstead, R.W., 1997, APJ, 478, 536

Pozzetti, L., Bruzual, A.G. \& Zamorani, G., 1996, MNRAS, 281, 953

Scalo, J.N., 1986, Fundamentals of Cosmic Physics, Vol. 11, p1

Shanks, T., Stevenson, P.R.F., Fong, R. \& McGillivray, H.T., 1984, MNRAS, 206, 767

Somerville, R., 1997, PhD thesis, University of California at Santa Cruz

Songaila, A., Cowie, L.L., Hu, E,M. \& Gardner, J.P., 1994, APJS, 94, 461

Szokoly, G.P., Subbarau, M.U., Connolly, A.J. \& Mobasher, B., 1998, APJ, 492, 452

Tinsley, B.M., 1980, APJ, 241, 41 

White, S.D.M. \& Rees, M.J., 1978, MNRAS, 183, 341

White, S.D.M. \& Frenk, C.S., 1991, APJ, 379, 52

Yoshii, Y. \& Takahara, F., 1988, APJ, 326, 1

Zepf, S., 1997, Nature, 390, 377

Zucca, E., Zamorani, G., Vettolani, G., Cappi, AS., Merighi, R., Mignoli, M.,
Stirpe, G.M., Macgillivray, H. et al., 1997, A\&A, 326, 477
\end {document}